\documentclass[twocolumn,english,prl,showpacs,superscriptaddress]{revtex4-1}

\usepackage[T1]{fontenc}
\usepackage[latin9]{inputenc}
\usepackage{amstext}
\usepackage{graphicx}
\usepackage{babel}
\usepackage{times}
\usepackage{amssymb}
\usepackage{amsfonts}
\usepackage{amsmath}

\begin{document}

\title{Theory of the Anomalous Tunnel Hall Effect at Ferromagnet-Semiconductor Junctions.\\ }

\author{T. Huong Dang}
\affiliation{Unit\'e Mixte de Physique, CNRS, Thales, Univ. Paris-Sud, Universit\'e Paris-Saclay, 91767 Palaiseau, France}
\author{D.~Quang To}
\affiliation{Laboratoire des Solides Irradi\'es, Ecole Polytechnique, CNRS UMR 7642, and CEA-DSM-IRAMIS, University Paris Saclay, 91128 Palaiseau cedex, France}
\author{E. Erina}
\affiliation{Laboratoire des Solides Irradi\'es, Ecole Polytechnique, CNRS UMR 7642, and CEA-DSM-IRAMIS, University Paris Saclay, 91128 Palaiseau cedex, France}
\author{T.~L. Hoai Nguyen}
\affiliation{Institute of Physics, VAST, 10 Daotan, Badinh, Hanoi, Vietnam.}
\author{V. Safarov}
\affiliation{Laboratoire de Physique de la Mati\`ere Condens\`ee, Ecole Polytechnique, University Paris Saclay, 91128 Palaiseau cedex, France}
\author{H. Jaffr\`es}
\affiliation{Unit\'e Mixte de Physique, CNRS, Thales, Univ. Paris-Sud, Universit\'e Paris-Saclay, 91767 Palaiseau, France}
\author{H.-J. Drouhin}
\affiliation{Laboratoire des Solides Irradi\'es, Ecole Polytechnique, CNRS UMR 7642, and CEA-DSM-IRAMIS, University Paris Saclay, 91128 Palaiseau cedex, France}

\begin{abstract}
We report on theoretical investigations of carrier scattering asymmetry at ferromagnet-semiconductor junctions. By an analytical $2\times 2$ spin model, we show that, when Dresselhaus interactions is included in the conduction band of III-V $T_d$ symmetry group semiconductors, the electrons may undergo a difference of transmission \textit{vs.} the sign of their incident parallel wavevector normal to the in-plane magnetization. This asymmetry is universally scaled by a unique function independent of the spin-orbit strength. This particular feature is reproduced by a multiband $\mathbf{k}\cdot \mathbf{p}$ tunneling transport model.  Astonishingly, the asymmetry of transmission persists in the valence band of semiconductors owing to the inner atomic spin-orbit strength and free of asymmetric potentials . We present multiband $14\times 14$ and $30\times 30$ $\mathbf{k}\cdot \mathbf{p}$ tunneling models together with tunneling transport perturbation calculations corroborating these results. Those demonstrate that a tunnel spin-current normal to the interface can generate a surface transverse charge current, the so-called Anomalous Tunnel Hall Effect.
\end{abstract}


\date{\today}
\maketitle

\section{I - Introduction}

Spinorbitronics is a science that uses the spin degree of freedom together with the spin-orbit interactions (SOI) to generate spin-currents without the need of a ferromagnetic material. Those spin-currents become essential in view to control the magnetization state of a nanomagnet, \textit{via} the spin-Hall effect of heavy material either semiconductor~\cite{dyakonov_a,dyakonov_b,sinova2004,awschalom2004,wunderlich2005} or metals~\cite{valenzuela2006,liu2012,miron2010,garello2013,rojas2014}. The interplay between particle spin and orbital motion is currently at the basis of a new family of effects like the Anomalous Tunnel Hall effect described by the generation of a charge current transverse to a tunneling spin-current~\cite{rylkov2017,jamet2006,dang2015,fabian2015} or the spin-galvanic effects~\cite{ganichev2016}. SOI at an interface with broken inversion symmetry may lead to the observation of Rashba-split states~\cite{bychkov1984,gambardella2011} which may be used to convert a flow of spin-current into a transverse charge current by inverse Edelstein effect ($IEE$)~\cite{rojas2013,lesne2016}. Recent magnetoresistance (MR) measurements with an unidirectional character (UMR) have been evidenced in metallic ~\cite{gambardella2015} and semiconductor bilayers~\cite{wunderlich2015} as well as with topological insulator (TI)~\cite{yasuda2016}. It was ascribed to the asymmetric scattering of electrons by magnons absorption-emission processes at ferromagnet-TI interfaces. It clearly reveals a novel symmetry in the field of MR explained by a certain carrier scattering asymmetry incorporated in the Boltzmann transport equation developed at the second order. An anisotropy of electronic diffusion on the Fermi surface along the current direction leads to a supplementary non-linear contribution of $MR$ proportional to the current.

In this article, as an extension to our previous work~\cite{dang2015}, we study unconventional quantum effects resulting in a giant transport asymmetry of carriers in semiconductor heterostructures, interfaces, tunnel barriers or quantum wells. Those are composed of ferromagnets and spin-orbit split electrodes made of semiconductors, \textit{e.~g.} III-V compounds, with magnetizations of opposite direction (AP). The symmetry of the structure allows a difference of transmission upon positive or negative incidence (that we will note $\xi=\pm \mathbf{k}_\parallel$) with respect to the reflection plane defined by the magnetization and the surface normal. However, this quantum process departs from the effect of a beam deviation by the action of the Lorentz force~\cite{alekseev2010} and, unlike spin-filtering effects~\cite{perel2003,perel2004}, the scattering asymmetry requires the simultaneous action of both in-plane and out-of plane spin-orbit fields in the case of Dresselhaus interaction in the condution band (CB). In a second part of the paper, we emphasize on the perturbation calculation techniques needed to understand this phenomena as well as the case of intrinsic SOI in the valence band (VB).

\section{II - Case of the conduction band of semiconductors of $T_d$ symmetry group.}

We first consider the Dresselhaus interactions~\cite{Dresselhaus}, $\widehat{H}_{D}=\left( \widehat{\gamma }\mathbf{\chi}\right) \cdot \widehat{\mathbf{\sigma}}$,

\begin{eqnarray}
\widehat{H}_{D}=\left(
\begin{array}{cc}
-\widetilde{\gamma }\xi ^{2}k & -i\gamma \xi k^{2} \\
i\gamma \xi k^{2} & \widetilde{\gamma}\xi ^{2}k%
\end{array}%
\right)
\end{eqnarray}
in the conduction band of a semiconductor junction made of two magnetic materials in the $AP$ state. We refer the structure to the $x, y, z$ cubic axes (unit vectors $\widehat{x},\widehat{y},\widehat{z}$) and assume that electron transport occurs along the $z$ axis, whereas the magnetization lies along $x$. One then introduces $(0,\xi, k)$ the electron wavevector; $\widehat{\mathbf{\sigma }}$ the Pauli operator, and $\mathbf{\chi}=\left[ 0,\xi k^{2},-\xi^{2}k\right]$ the D'yakonov-Perel' (DP) internal field responsible for the spin splitting~\cite{DyaPer,Dresselhaus}. One introduces the tensor $\widehat{\gamma }=\left( \gamma_{i}\delta _{ij}\right) $ which characterizes the DP-field strength, with $\gamma _{x}=\gamma _{y}=\gamma $, $\gamma _{z}=\widetilde{\gamma }$, and $\delta _{ij}$ the Kronecker symbol. We consider the two cases $ \widetilde{\gamma }=\gamma $ and $\widetilde{\gamma }=0$ on switching off the $\xi ^{2}$ perturbation.

\begin{figure}
\includegraphics[height=12cm]{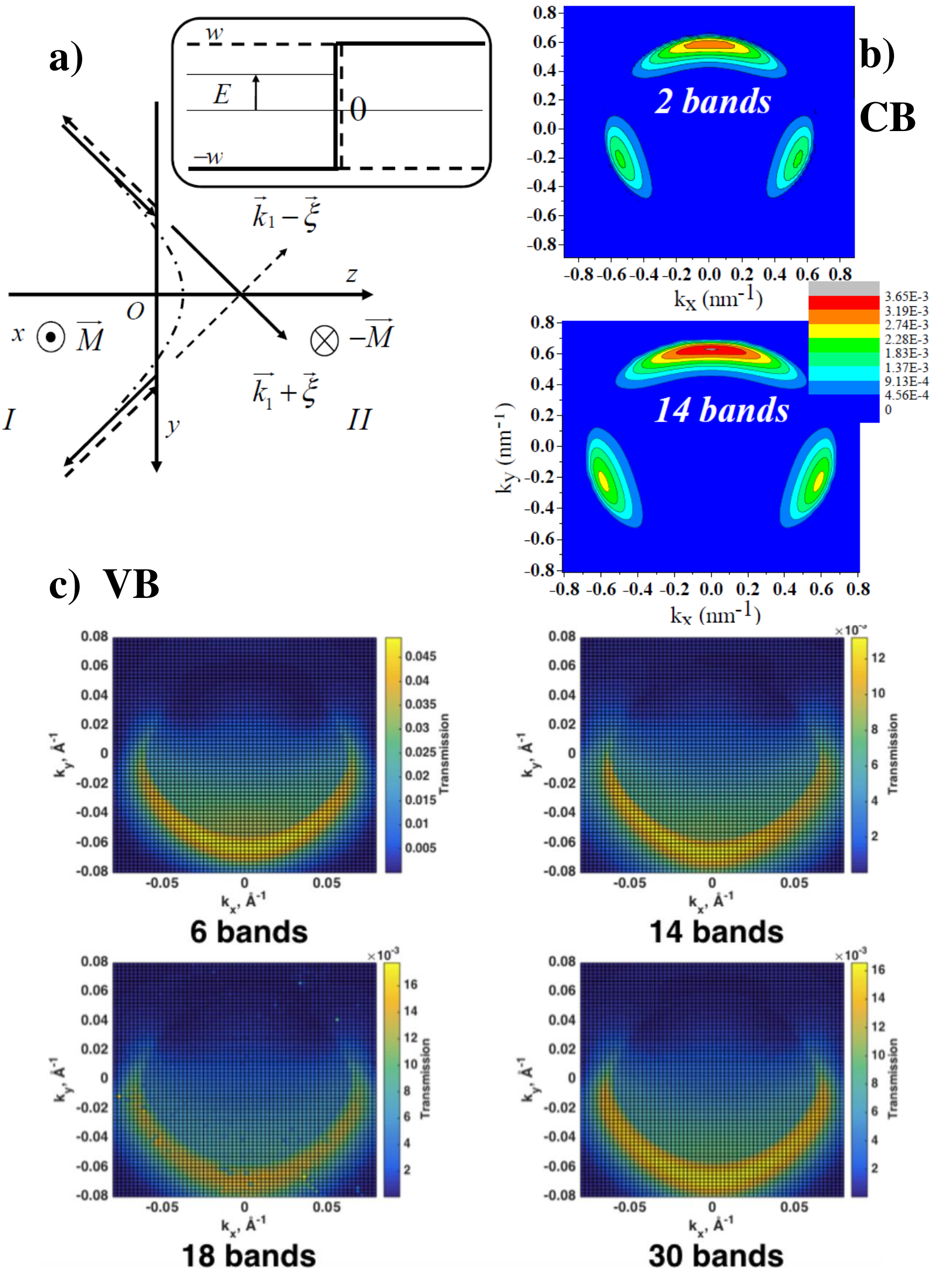}
\caption{Scheme of the transmission process at an exchange-SOI step with $AP$ magnetizations. $k_{1}$ are the propagative wavevector whereas $\pm \xi$ are the parallel ones. Carriers with $+\xi $ in-plane wavevector component are more easily transmitted than those carrying $-\xi$. (Top right inset): Energy profile of the exchange step;  $E$ is the longitudinal kinetic energy and $2w$ is the exchange splitting. b) Asymmetry in the CB: 2D-map of the transmission coefficient $T$ in a $2\times 2$ and $14\times 14$ $\mathbf{k}\cdot \mathbf{p}$ band model. $\protect\gamma =-24$~eV~\AA $^{\text{3}}$. c) Asymmetry in the VB: 2D-map of the transmission coefficient $T$ of a tunnel junction within a $6\times 6$, $14\times 14$, $18\times 18$ and $30\times 30$ $\mathbf{k}\cdot \mathbf{p}$ band models. In both CB and VB, the exchange strength is $0.3$~eV and the \textit{total} kinetic energy $\mathcal{E}=0.23$ eV. For VB, the barrier thickness is 3 nm. The respective 14-band and 30-band parameters are taken from Refs.~[\cite{jancu2005}] and [\cite{richard}]}\label{fig1}
\end{figure}

We study the transmission asymmetry when the wavevector component along $y$ is changed from $\xi $ to $-\xi $. Electrons are injected from the first conduction band of material $I$ to the left ($\epsilon =1$) into the first conduction band of material $II$ to the right ($\epsilon =-1$). Then, the relevant $2\times 2$ Hamiltonians respectively write:
\begin{eqnarray}
\widehat{H}_{I,II} &=&\gamma _{c}\left( k^{2}+\xi ^{2}\right) \widehat{I}+w\mathbf{m}\cdot \widehat{\mathbf{\sigma }}+\widehat{H}_{D}  \nonumber \\
&=&\left(
\begin{array}{cc}
\gamma _{c}(k^{2}+\xi ^{2})-\widetilde{\gamma }\xi ^{2}k & -i\gamma \xi k^{2}+\epsilon w \\
i\gamma \xi k^{2}+\epsilon w & \gamma _{c}(k^{2}+\xi ^{2})+\widetilde{\gamma
}\xi ^{2}k%
\end{array}%
\right)   \label{H}
\end{eqnarray}
where $\widehat{I}$ is the identity matrix,\ $\gamma _{c}$ the conduction effective mass, $\mathbf{m}$ is the unit magnetization, $2w$ the exchange strength.

\vspace{0.1in}

\emph{Transmission from quantum boundary conditions (quantum wavefunction matching).}

\vspace{0.1in}

The two energies in the electrodes are given by $\mathcal{E}_{1}=$ $\gamma _{c}\left(k_{1}^{2}+\xi ^{2}\right) -w$ and $\mathcal{E}_{2}=\gamma _{c}(k_{2}^{2}+\xi^{2})+w$, where $k_{1}$ $\left(k_{2}-\text{pure imaginary}-\right)$ is the $z$-component of the wavevector in the lower (upper) subband. The two eigenvectors  write:
\begin{eqnarray}
\mathbf{u}_{\epsilon ,1}\left( \xi ,k_{1}\right)  &=&\left[ 1-2\epsilon i\mu k_{1}^{2},-\epsilon \left( 1-2\widetilde{\mu }\xi k_{1}\right) \right] /\sqrt{2}\text{,}  \label{u1} \\
\mathbf{u}_{\epsilon ,2}\left( \xi ,k_{2}\right)  &=&\left[ 1-2\epsilon i\mu k_{2}^{2},\epsilon \left( 1+2\widetilde{\mu }\xi k_{2}\right) \right] /\sqrt{2}\text{,}  \label{u}
\end{eqnarray}%
where $\mu =\gamma \xi /(2w)$ and $\widetilde{\mu }=\widetilde{\gamma }\xi /(2w)$. Note that the form of the eigenvectors does not foresee any tunneling transmission asymmetry in usual tunneling models~\cite{dang2015} based on the density of states~\cite{julliere1975,slonczewski1989}. The asymmetry arises from a full-quantum treatment discussed in terms of chirality. Because $k_{\parallel}$ is conserved, we are dealing with states with the same \textit{longitudinal} kinetic energy $E$ along $z$ and a total kinetic energy $\mathcal{E}=E+\gamma _{c}\xi ^{2}$. The boundary conditions are the continuity of the wavefunction and of the current wave $\widehat{J}\Psi_{I,II}=\left(1/\hbar \right) (\partial \widehat{H}_{I,II}/\partial k) \Psi_{I,II}$ because $\widehat{H}_{I,II}$ contains no more than quadratic $k$ terms~\cite{petukhov2002,elsen2007bis,Hoai2009,Drouhin2011,BottPRB2012}.

The transmission of a pure up-spin incident electron into a pure down-spin state is only possible under oblique incidence \textit{via} SOI which introduces off-diagonal matrix elements. The spin-orbit field is also responsible for a discontinuity of the spin-current between incident (\textit{inc}) and transmitted (\textit{trans.}) waves. Moreover, a non-vanishing diagonal part of SOI is necessary to obtain a non-zero asymmetry although the $z$ component of the DP field along $z$ does not depend on the sign of $k_{\parallel}$~\cite{dang2015}. Then, from now on, we take $\widetilde{\gamma }=\gamma $. The wavevector $k_{1}$ in the lower subband has to be real so that we can define $K=k_{1}>0$. We introduce the parameter $\lambda>0$ with $k_{2}=i\lambda K$, the reduced longitudinal energy $\eta=E/w=\left( 1-\lambda^{2}\right) /\left( 1+\lambda ^{2}\right)$, and the incidence parameter $t=\xi /K$. One finally obtains the transmission $T\left(t,\eta \right)$ and its average $\overline{T}$ upon $\pm \xi$ incidence $t$ and asymmetry $\mathcal{A}$:

\small

\begin{eqnarray}
\overline{T}\left(|t|,\eta\right)=\frac{\gamma ^{2}}{\gamma _{c}^{3}}wt^{2}\left(1+\eta \right) ^{2}\left[4\eta ^{2}\left(1-\eta \right)+t^2 (1+\eta)(2\eta -1)^{2} \right] \text{,}\\
\mathcal{A}\left(t,\eta \right) \ =\frac{4t \eta \sqrt{1-\eta ^{2}}\left(2\eta -1\right) }{4\eta ^{2}\left( 1-\eta \right) +t^{2}\left( 1+\eta \right) \left( 2\eta -1\right) ^{2}}\text{.}
\label{TA}
\end{eqnarray}\normalsize
where $\mathcal{A}=\left[T\left(t,\eta \right)-T\left(-t,\eta \right)\right]/\left[T\left(t,\eta \right)+T\left(-t,\eta \right)\right]$, emphasizing the increase of $T\left( t,\eta \right)$ with $t$ and $\gamma$. The analytical asymmetry $\mathcal{A}$ is plotted in Fig.~2a for several values of $t$ (full lines), where the symbols refer to the $2\times 2$ numerical calculations, showing an excellent agreement. It is a remarkable result that $\mathcal{A}\left(t,\eta\right) $ does not depend either on the material parameters or on the sign of $\gamma$, thus conferring to $\mathcal{A}$ a universal character. Reversing the magnetization (changing $w$ into $-w$) makes transport occur in the $k_{2}$ channel leading to a change of $\mathcal{A}\left(t,\eta \right)\ $ into $-\mathcal{A}\left( t,\eta\right)$. Our convention is that $\mathcal{A}$ is positive, at small energy $\eta$ (or averaged over the energy band) when $\left( \mathbf{m},\mathbf{\xi},\mathbf{k}\right) $ forms a direct frame and negative otherwise. Another striking feature is that an arbitrarily small perturbation is able to produce a 100\% transport asymmetry \textit{i.e.}, a total quenching of transmission in the CB. Fig.~1b display the 2-dimensional map of the electron transmission at a given total energy in the reciprocal space calculated using both a $2\times 2$ effective Hamiltonian and a full $14\times 14$ band $\mathbf{k}\cdot \mathbf{p}$ treatment involving odd-potential coupling terms $P^{^{\prime }}$ and $\Delta ^{^{\prime }}$~\cite{jancu2005,cardona1988,pfeffer1991}. These calculations are based on the multiband transfer matrix technique detailed in Refs.~\cite{petukhov2002, elsen2007bis}.

\vspace{0.1in}

\emph{Transverse Surface Currents.}

\vspace{0.1in}

The transmitted current summed, $\mathbf{J}\left[t,\eta\right]=\mathbf{J}_{\xi}\left[\Psi_{II}\left(z\right) \right] +\mathbf{J}_{-\xi}\left[\Psi_{II}\left(z\right)\right]$, originates from incident waves of equal amplitude and opposite $k_{\parallel}$. To the lowest order in $\gamma$, we find
\begin{equation}
\mathbf{J}_{y,z}\left[ t,\eta\right]=\frac{4\left(\gamma_{c}w\right)
^{1/2}}{\hbar}\left(1+\eta\right)^{1/2}T\left(t,\eta\right)  \left[
\mathcal{A}\left(t,\eta\right)  t\widehat{y}\mathbf{+}\widehat{z}\right]
\label{current}
\end{equation}
A non-zero $\mathcal{A}$ gives rise to a transverse carrier momentum and then to a tunneling surface current (per unit length) $j_y=J_y \times \ell_{mfp}$ ($\ell_{mfp}$ is the electron mean free path), of the form $\mathbf{J}_{C}=\mathbf{m}\times \mathbf{J}_{S,z}$ ($C$ for current and $S$ for spin-current), leading to a potentially large Anomalous Tunnel Hall Effect ($ATHE$). The ratio of the surface transverse to the longitudinal current density, $j_{_{y}}\left[t,\eta\right]/J_{z}\left[t,\eta\right]=t\mathcal{A}\left(t,\eta\right)\ell_{mfp}$, leads to the $ATHE$ length, or $ATHE$ angles~\cite{dang2015}, in the spirit of the work dealing with IEE phenomenon~\cite{rojas2013,lesne2016}.

\section{III - Case of the CB: Perturbation Calculations involving SOI.}

Using advanced perturbation procedures, one may give a general expression for the change of the transmission amplitude $\delta t^{\sigma \sigma^{\prime}}$ of a propagative spin-$\uparrow$ wave from the left transmitted into a propagative spin-$\downarrow$ wave to the right, after having experienced a SOI potential $V^{\sigma\sigma^{\prime}}$ (spin-flip) in a confined region of space. The calculation is based on Ref.~\cite{alekseev2014} and we will demonstrate \textit{in fine} that,

\begin{eqnarray}
\delta \Psi _{in}^{0\sigma}(z)=\int_{0}^{a}G_0^{\sigma\sigma}(z,z^{\prime})V^{\sigma \sigma ^{\prime}}(z^{\prime })\Psi _{in}^{0\sigma ^{\prime }}(z^{\prime })dz^{\prime},
\label{equ:tsf}
\end{eqnarray}
and from the expression of $G_0$, that $t^{\uparrow \downarrow}$ may be written:

\begin{eqnarray}
\delta t^{\sigma \sigma^{\prime}}=\frac{-im^{\ast }}{\hbar ^{2}k}\int_{0}^{a}\Psi _{out}^{0\sigma }(z^{\prime })V^{\sigma \sigma ^{\prime}}(z^{\prime })\Psi _{in}^{0\sigma ^{\prime }}(z^{\prime })dz^{\prime},
\label{equ:tspf}
\end{eqnarray}
where (\textit{in}) and (\textit{out}) refer respectively to the unperturbed \textit{incoming} wave at left and \textit{outgoing} wave at right~\cite{alekseev2014}. $G_0^{\sigma\sigma}$ is the (spin-diagonal) Green's function (GF) to consider and we are searching for. Such perturbative scattering approach has hardly been employed to investigate the role of the evanescent waves in transport like investigated here. The method is particularly suitable for the case of non-degenerate orbital systems but however could be applied, in a future work, to the case of the valence band (VB). We consider the Green's function (GF) $G_0$ of an hamiltonian system $\hat{H}_{0}=\left(\frac{\hbar ^{2}}{2}\right) \frac{\partial }{\partial z}\frac{1}{m^{\ast }(z)}\frac{\partial }{\partial z}-U(z)$ in a homogenous potential $U_{1}$ for $z<0$, and $U_{2}$ for $z>0$ satisfying:

\begin{eqnarray}
\left( \mathcal{E}-\hat{H}_{0}\right)
G_{0}^{\sigma\sigma}(z,z^{\prime }) &=&\delta \left(z-z^{\prime}\right),
\label{Scattering 8}
\end{eqnarray}%

\emph{Green's function without orbital degeneracy.}

\vspace{0.1in}

The strategy to find the GF is then \textit{i)} to find two different ground states $\left\{\Psi _{L}^{0,\uparrow\downarrow},\Psi_{R}^{0,\uparrow\downarrow}\right\} $ of the homogenous Schr\"{o}dinger equation $\left(\mathcal{E}-\hat{H}_{0}\right) \Psi =0$ ($L$ for left and $R$ for right whith characteristic wavevector $k_{I}$ and $k_{II}$), \textit{ii)} to find the relevant linear combinations of $\Psi _{L}^{0}$ and $\Psi_{R}^{0}$ that make $y_{1}$ and $y_{2}$ solution of the equation $\left(\mathcal{E}-\hat{H}_{0}\right) y=0$ finite at $z=-\infty$ [$y_{1}(z)$] and $z=+\infty$ [$y_{2}(z)$] depending on the use of the \textit{retarded} or \textit{advanced} quantities, and \textit{iii)} to define the correct GF by making use of:
\begin{equation}
G_{0}^{\sigma\sigma}(z,z^{\prime })=\left\{
\begin{array}{c}
\frac{y_{1}^{\sigma}(z)y_{2}^{\sigma}(z^{\prime })}{W(y_{1},y_{2})} -\infty <z<z^{\prime }<+\infty \\
\frac{y_{1}^{\sigma}(z^{\prime})y_{2}^{\sigma}(z)}{W(y_{1},y_{2})} -\infty <z^{\prime }<z<+\infty
\end{array}
\right. ,
\end{equation}%
where $W(y_1,y_2)=\frac{\hbar ^{2}}{2m^{\ast}}\left[y_{1}(z^{\prime})\frac{\partial y_{2}(z^{\prime})}{\partial z^{\prime }}-\frac{\partial y_{1}(z^{\prime })}{\partial z^{\prime }}y_{2}(z^{\prime })\right]$ is the Wronskian. The homogenous Schr\"{o}dinger equation, $\left( \mathcal{E}-\hat{H}_{0}\right) \Psi=0$, admits the solutions:
\begin{eqnarray}
\Psi _{L}^{0,\sigma}=\left(e^{ik_{I}^\sigma z_{<}}+r_{L}^{\sigma}e^{-ik_{I}^\sigma z_{<}}+t_{L}^{\sigma}e^{ik_{II}^\sigma z_{>}}\right)|\sigma> \nonumber\\
\Psi _{R}^{0,\sigma}=\left(e^{-ik_{II}^\sigma z_{>}}+r_{R}^{\sigma}e^{ik_{II}^\sigma z_{>}}+t_{R}^{\sigma}e^{ik_{I}^\sigma z_{<}}\right)|\sigma>
\label{eqn:waves}
\end{eqnarray}%
where $z_{<}$ and $z_{>}$ stand for $z<0$ and $z>0$. If we chose $y_1=\Psi _{R}^{0,\sigma}$ and $y_2=\Psi _{L}^{0,\sigma}$, Eq.~\ref{Scattering 8} admits a particular solution:

\scriptsize

\begin{equation}
G_{0}^{\sigma\sigma}(z,z^{\prime })=\frac{\Psi _{R}^{0,\sigma}(z^{\prime})\Psi _{L}^{0,\sigma}(z)\Theta (z-z^{\prime})+\Psi _{R}^{0,\sigma}(z)\Psi _{L}^{0,\sigma}(z^{\prime})\Theta (z^{\prime}-z)}{W(\Psi _{R}^{0},\Psi _{L}^{0})},
\label{interface Green}
\end{equation}

\normalsize

On the assumption of a same effective mass, the Wronskian $W=\frac{\hbar ^{2}}{2m^{\ast}}\frac{4ik_{I}^\sigma k_{II}^\sigma}{k_{II}^\sigma+k_{I}^\sigma}$ is a constant ($\partial W/\partial z^{\prime}=0$) and we recover the \textit{retarded} GF introduced in Ref.~\cite{Stewart_2003} according to:

\small

\begin{eqnarray}
G_{0}(z_{<},z^{\prime}_{>})=\frac{2m^{\ast}}{\hbar ^{2}}\frac{t_{L}}{2ik_{I}^\sigma}e^{-ik_{I}^\sigma z}e^{ik_{II}^\sigma z^{\prime }}, \nonumber\\
G_{0}(z_{>},z^{\prime}_{<})=\frac{2m^{\ast}}{\hbar ^{2}}\frac{t_{R}}{2ik_{II}^\sigma}e^{-ik_{I}^\sigma z^{\prime }}e^{ik_{II}^\sigma z}, \nonumber\\
G_{0}(z_{<},z^{\prime}_{<})=\frac{2m^{\ast}}{\hbar ^{2}}\frac{1}{2ik_{1}^\sigma}\left[e^{ik_{I}^\sigma \left\vert z-z^{\prime}\right\vert}+r_{L}^\sigma e^{-ik_{I}^\sigma(z+z^{\prime})}\right], \nonumber\\
G_{0}(z_{>},z^{\prime}_{>})=\frac{2m^{\ast}}{\hbar ^{2}}\frac{1}{2ik_{II}^\sigma}\left[e^{ik_{II}^\sigma\left\vert z-z^{\prime}\right\vert}+r_{R}^\sigma e^{-ik_{II}^\sigma(z+z^{\prime})}\right]
\end{eqnarray}\normalsize
that we will use henceforth.

\begin{figure}[tbp]
\includegraphics[height=10cm]{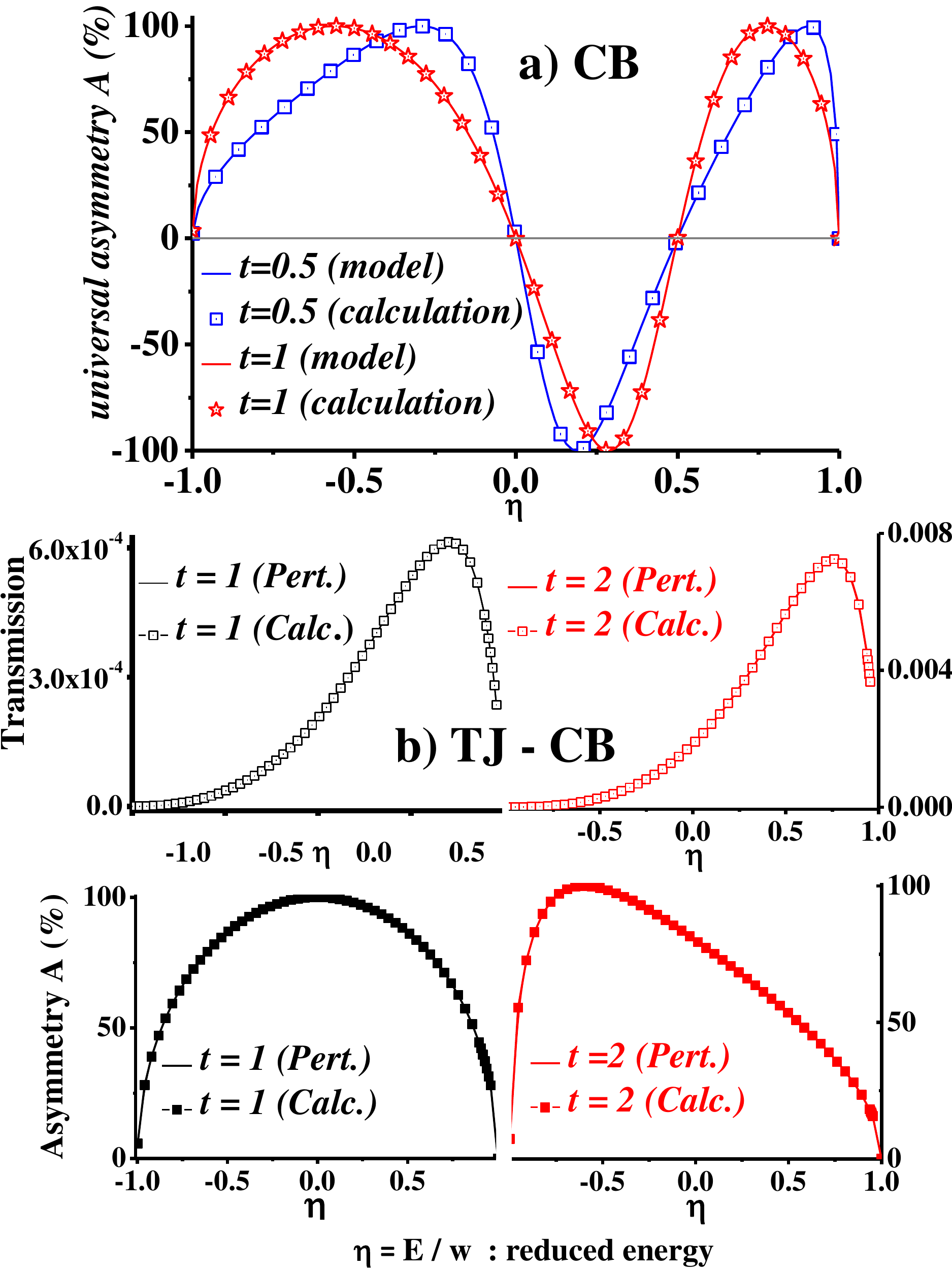}
\caption{(a) Universal asymmetry coefficient $\mathcal{A}$ \textit{vs.} reduced energy $\protect\eta =E/w$ obtained for an exchange-step with different values of $t=\protect\xi /K$\ [$t=0.01$ (black; circles), $t=0.5$ (blue; squares), $t=1$ (red; stars), and $t=2$ (purple; triangles) by $2$-band analytical (full line) and numerical (symbols) calculations. (b) Transmission coefficients and asymmetry coefficient $\mathcal{A}$ \textit{vs.} reduced energy $\protect\eta =E/w$ obtained for a 3~nm tunnel junction (TJ) with different values of $t=\xi/K$\ [$t=1$ (black), $t=2$ (red), by perturbative scattering (\textit{pert.}: full lines) method and numerical $\mathbf{k}\cdot \mathbf{p}$ calculations (\textit{Calc.}: symbols).}
\end{figure}

\subsection{Case of the exchange-step in the CB: perturbation calculations. }

We revisit here the issue and results of section II (exchange-step) with $\hat{H}_{0}$, eigenvectors and eigenvalues given in section II. We recall that the current is along $\hat{z}$ and magnetizations along $\hat{x}$. The incident wavevector is $\mathbf{k}=\left(0,\xi,k\right)$.

\vspace{0.1in}

\emph{Reflection, transmission and perturbating SOI potential}

\vspace{0.1in}

We consider the electron transmission within an energy window in the exchange step, $-w<\mathcal{E}<w,$ where the transmission asymmetry takes place, so that $k_{I}^\uparrow$ and $k_{II}^\downarrow$ are real whereas $k_{I}^\downarrow$ and and $k_{II}^\uparrow$ are pure imaginary. In the right contact, the spin $\uparrow$ state admits a pure propagative character whereas the spin $\downarrow $ state is purely evanescent. It is then quite convenient to define $k_{I}^\uparrow=k_{II}^\downarrow=k_1$ and  $k_{I}^\downarrow=k_{II}^\uparrow=ik_2$.  The two solutions of the homogeneous Schr\"{o}dinger equation, $\Psi_{L}^{0,\uparrow}$, and $\Psi _{R}^{0,\downarrow}$ are given by Eq.[~\ref{eqn:waves}] with reflection, $r_{L\uparrow }=r_{R\downarrow}=\frac{k_{1}-ik_{2}}{k_{1}+ik_{2}}$, and transmission, $t_{L\uparrow}=t_{R\downarrow}=\frac{2k_{1}}{k_{1}+ik_{2}}$ ($t_{R\uparrow}=t_{L\downarrow }=\frac{2k_{2}}{k_{2}-ik_{1}}$), amplitudes found via the matching conditions at $z=0$. This allows possible transmission from propagative to evanescent states ($t_{R\downarrow}$ and $t_{L\uparrow }$) and vice-versa ($t_{L\uparrow }$ and $t_{L\downarrow }$).

The SOI, $\hat{H}_{D}^{\sigma\sigma^{\prime}}$, is then introduced as a perturbation potential according to:

\small

\begin{eqnarray}
\hat{H}_{D} &=&\ -\left[\frac{\xi ^{2}\sigma _{z}}{2}\left(\gamma k+k^{+}\gamma (z)\right)-\frac{\xi \sigma _{y}}{2}\left( \gamma k^{2}+\left( k^{+}\right)^{2}\gamma \right) \right] \nonumber\\
&=&\frac{i\xi ^{2}\sigma _{z}}{2}\left(\gamma \overrightarrow{\frac{\partial}{\partial z}}-\frac{\overleftarrow{\partial}}{\partial z}\gamma \right) -\frac{\xi \sigma _{y}}{2}\left(\gamma \overrightarrow{\frac{\partial ^{2}}{\partial z^{2}}}+\frac{\overleftarrow{\partial ^{2}}}{\partial z^{2}}\gamma \right),
\end{eqnarray}\normalsize
with $\gamma=\gamma(z)$. $\hat{H}_{D}$ acquires a pure non-diagonal form like:

\scriptsize
\begin{eqnarray}
V^{\uparrow \downarrow } &=&\langle \uparrow |\hat{H}_{D}|\downarrow \rangle, \nonumber \\
&=&\langle \uparrow |\left\{\frac{i\xi ^{2}\sigma _{z}}{2}\left( \gamma \overrightarrow{\frac{\partial }{\partial z}}-\frac{\overleftarrow{\partial}}{\partial z}\gamma \right)-\frac{\xi \sigma _{y}}{2}\left(\gamma \overrightarrow{\frac{\partial ^{2}}{\partial z^{2}}}+\frac{\overleftarrow{\partial ^{2}}}{\partial z^{2}}\gamma\right) \right\} |\downarrow
\rangle, \nonumber\\
&=&\left(\frac{i\xi ^{2}}{2}\gamma \overrightarrow{\frac{\partial }{\partial z}}-\frac{i\xi }{2}\gamma \overrightarrow{\frac{\partial ^{2}}{\partial z^{2}}}\right)-\left(\frac{i\xi ^{2}}{2}\frac{\overleftarrow{\partial}}{\partial z}\gamma +\frac{i\xi }{2}\frac{\overleftarrow{\partial ^{2}}}{\partial z^{2}}\gamma \right),
\end{eqnarray}\normalsize
and
\scriptsize
\begin{eqnarray}
V^{\downarrow \uparrow } 
&=&\left(\frac{i\xi ^{2}}{2}\gamma \overrightarrow{\frac{\partial }{\partial z}}+\frac{i\xi }{2}\gamma \overrightarrow{\frac{\partial ^{2}}{\partial z^{2}}}\right)-\left( \frac{i\xi^{2}}{2}\frac{\overleftarrow{\partial }}{\partial z}\gamma-\frac{i\xi }{2}\frac{\overleftarrow{\partial^{2}}}{\partial z^{2}}\gamma \right)
\end{eqnarray}\normalsize

From Eq.~[\ref{equ:tsf}] and $W=i\frac{\hbar ^{2}k_{1}}{m^{\ast }}t_{R\uparrow}$, the correction to the amplitude of transmission is:
\small
\begin{equation}
\begin{split}
\delta t^{\uparrow \downarrow} =-\frac{m^{\ast}}{i\hbar ^{2}k_{1}}\int_{-\infty }^{+\infty }\Psi_{R}^{0\uparrow }(z^{\prime})\{-\frac{i\xi ^{2}}{2}\gamma \frac{\partial \Psi _{L}^{\downarrow 0}(z^{\prime})}{\partial z}+\frac{i\xi }{2}\gamma \frac{\partial ^{2}\Psi _{L}^{\downarrow 0}(z^{\prime })}{\partial z^{2}}- \\
-\frac{i\xi ^{2}}{2}\gamma \frac{\partial \Psi _{R}^{0\uparrow}(z^{\prime})}{\partial z}+\frac{i\xi}{2}\gamma \frac{\partial ^{2}\Psi_{R}^{0\uparrow }(z^{\prime})}{\partial z^{2}}\}\Psi _{L}^{\downarrow 0}(z^{\prime}) dz^{\prime}.
\end{split}
\label{Eq:trans}
\end{equation}\normalsize

We are now going to calculate the properties of the carrier transmission $\mathcal{A}$ for the different SOI configurations: at left, at right, and SOI\ in both contacts for an incoming left electron.

\vspace{0.1in}

\emph{SOI at left for electrons incoming from left.}

\vspace{0.1in}

We first note that the zeroth-order transmission coefficient, $t^{\uparrow\downarrow}_0=0$ is zero without spin-mixing interactions. Then, from Eq.~\ref{Eq:trans}, the transmission amplitude, $t^{\uparrow\downarrow}_L$,  with SOI at left is:

\scriptsize

\begin{equation}
\begin{split}
\delta t^{\uparrow \downarrow }=-\frac{m^{\ast}}{i\hbar ^{2}k_{1}}\int_{-\infty }^{0}\Psi_{R}^{0\uparrow }(z^{\prime})\{-\frac{i\xi ^{2}}{2}\gamma \frac{\partial \Psi _{L}^{\downarrow 0}(z^{\prime})}{\partial z}+\frac{i\xi}{2}\gamma \frac{\partial ^{2}\Psi _{L}^{\downarrow 0}(z^{\prime })}{\partial z^{2}}-\\
-\frac{i\xi ^{2}}{2}\gamma \frac{\partial \Psi _{R}^{0\uparrow }(z^{\prime})}{\partial z}+\frac{i\xi}{2}\gamma \frac{\partial ^{2}\Psi_{R}^{0\uparrow }(z^{\prime})}{\partial z^{2}}\}\Psi _{L}^{\downarrow 0}(z^{\prime})dz^{\prime}.
\end{split}
\end{equation}\normalsize

By considering $k_{1}=K$ (incoming propagative wavevector) and $k_{2}=\lambda K$ (imaginary transmitted wavevector), one then obtains:
\begin{equation}
t^{\uparrow\downarrow}_L=-\frac{1}{2w}\frac{\gamma \xi K^{2}}{\left( 1+i\lambda \right) ^{2}}\left\{ \frac{\xi }{K}(3\lambda ^{2}-1)+2\lambda \left( \lambda
^{2}-1\right) \right\}
\end{equation}

\vspace{0.05in}

\emph{SOI at right for electrons incoming from left.}

\vspace{0.05in}

The transmission changes from the previous case by changing the integral from $\int_{-\infty}^{0}$ into $\int_{0}^{+\infty}$ giving $t^{\uparrow\downarrow}_R=t^{\uparrow\downarrow}_L$.

\vspace{0.05in}

\emph{SOI at left and right side for electrons incoming from left.}

\vspace{0.05in}

With SOI in the whole space (left and right), we find from Eq.~[\ref{Eq:trans}] the transmission amplitude $t^{\uparrow\downarrow}$:

\scriptsize
\begin{eqnarray}
t^{\uparrow\downarrow}=-\left(\frac{\gamma \xi K^{2}}{w}\right)\frac{\frac{\xi }{K}(3\lambda ^{2}-1)+2\lambda \left( \lambda ^{2}-1\right)
}{\left(1+i\lambda \right) ^{2}}
\end{eqnarray}\normalsize

From $T(t,\eta)=|t^{\uparrow\downarrow}|^2$, we recover transmission $T(t,\eta)$ and asymmetry $\mathcal{A}$ derived from the application of the pure matching conditions (Eqs.~[\ref{TA}]). This proves the power of this perturbation methodology involving mixed propagative and evanescent electronic states.

\subsection{Case of a tunnel junction in the CB: perturbation calculations.}

\begin{figure}[tbp]
\includegraphics[height=6cm]{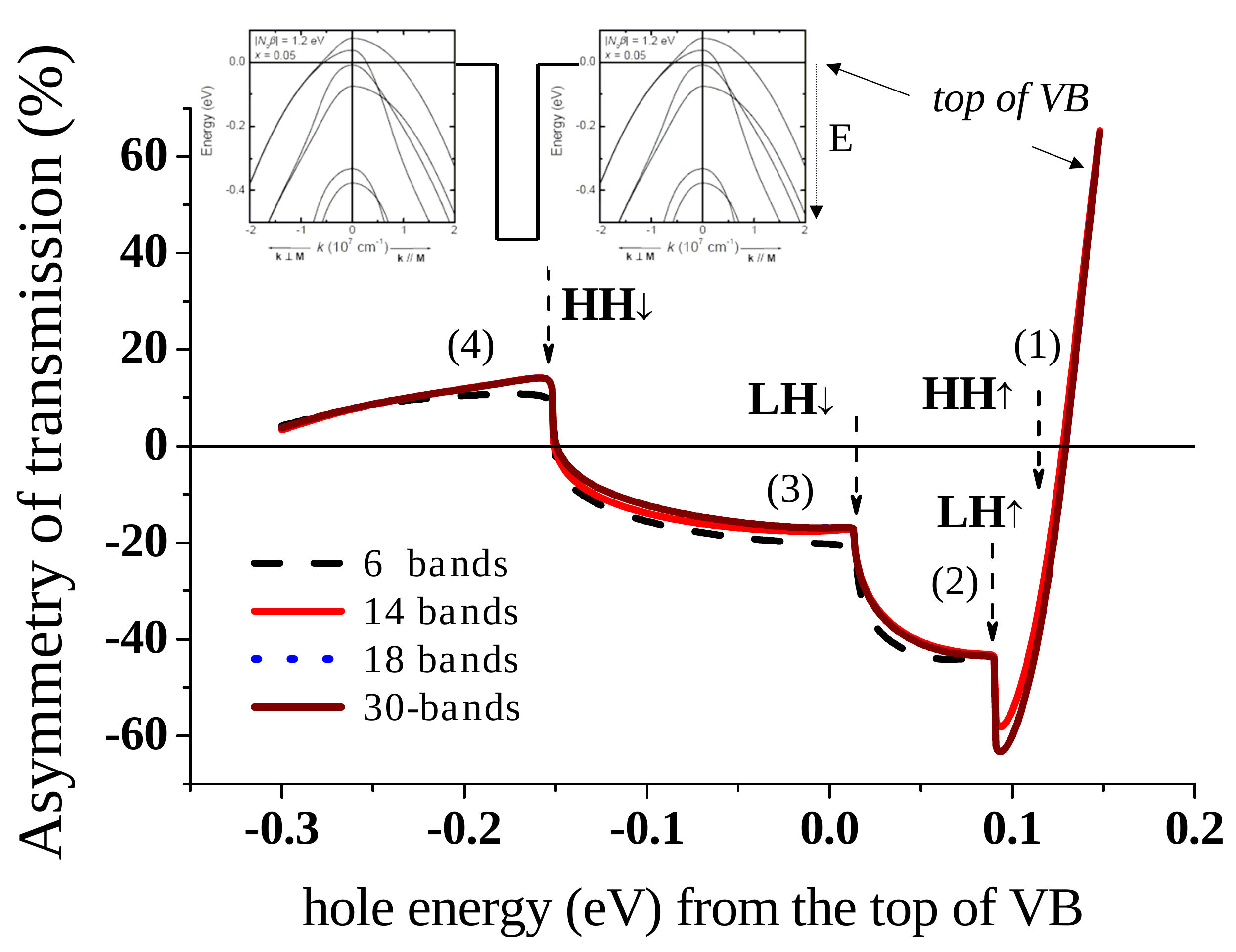}
\caption{Asymmetry coefficient $\mathcal{A}$ \textit{vs.} reduced energy calculated in the VB of GaMnAs/GaAs/GaMnAs 3~nm thick tunnel junction with an exchange strength of 0.3~eV and $k_\parallel=0.05~nm^{-1}$. The barrier height is -0.55~eV.}
\end{figure}

\normalsize

We focus now on the case of a tunnel junction, of thickness $a$, made of two ferromagnetic contacts (in the AP state) and separated by a thin semiconductor belonging to the T$_{\text{d }}$-symmetry. The contacts are free of SOI. The incident energy in the CB lies in the range of the exchange step, $-w<\mathcal{E}<w$, with a single incident propagative wave of a pure spin $\uparrow$ character. However, the electrons may scatter, now, at the two different interfaces of the junction and this makes the problem generally different from the previous treatment. One then considers a particular value for the barrier height equal to the exchange potential, $V_{0}=\left\vert w\right\vert,$ so as to prevent any back and forth scattering. The calculation of the most general shape of the GF is given in Ref.~\cite{de Aguiar 1993}. To the first order of perturbation, the transmission, $\delta t^{\downarrow \uparrow}$, now equals:

\scriptsize
\begin{equation}
\begin{split}
\delta t^{\uparrow \downarrow}=\frac{m^{\ast}}{i\hbar ^{2}k_{1}}\int_{0}^{a}\Psi _{L}^{0\uparrow }(z^{\prime })\{-\frac{i\gamma \xi ^{2}}{2}\frac{\partial \Psi _{R}^{\downarrow 0}(z^{\prime })}{\partial z}+\frac{i\gamma\xi }{2}\frac{\partial ^{2}\Psi _{R}^{\downarrow 0}(z^{\prime})}{\partial z^{2}}-\\
-\frac{i\gamma \xi ^{2}}{2}\frac{\partial \Psi _{L}^{0\uparrow }(z^{\prime })}{\partial z}+\frac{i\gamma \xi}{2}\frac{\partial ^{2}\Psi _{L}^{0\uparrow}(z^{\prime })}{\partial z^{2}}\}\Psi _{R}^{\downarrow 0}(z^{\prime})dz^{\prime},
\end{split}
\end{equation}
\normalsize

The coefficient of the wave functions $\Psi _{R}^{0\downarrow },$ and $\Psi_{L}^{0\uparrow }$, without SOI, are found from the relevant matching condition in a similar way to the case of the exchange step to give:

\begin{eqnarray}
\Psi _{L}^{0\uparrow}=t_{L\uparrow}e^{-k_{2} z}=\frac{2k_{1}}{k_{1}+ik_{2}}e^{-k_{2} z},\text{ for }z>0 \nonumber\\
\Psi _{R}^{0\downarrow }=t_{L\uparrow }e^{k_{2}(z-a)}=\frac{2k_{1}}{k_{1}+ik_{2}}e^{k_{2}(z-a)}\text{, for }z<a
\label{left wf in barrier}
\end{eqnarray}
Detailed calculations give out the transmission coefficient we are searching for:

\begin{equation}
\delta t^{\uparrow \downarrow }=\frac{e^{-k_{2} a}}{\gamma _{c}}\frac{2\gamma \xi k_{2} k_{1} a}{\left( k_{1}+ik_{2}\right)^{2}}(\xi +k_{2}).
\label{transmission amplitude in tunnel junction result}
\end{equation}
where we remind that $a$ is the barrier thickness.

Without SOI perturbation, the transmission coefficient is also zero in the situation of pure spin states, and consequently, $T^{\uparrow \downarrow }=\left\vert \delta t^{\uparrow \downarrow}\right\vert ^{2}.$ If one defines again the incidence parameter $t=\tan \theta =\xi /K$ for and $\eta =\frac{1-\lambda ^{2}}{1+\lambda ^{2}}=\frac{\mathcal{E}}{w}$ the reduced incident kinetic energy, we find the asymmetry of transmission for the tunnel barrier like:

\begin{equation}
\mathcal{A}=\frac{\left\vert \xi +k_{2}\right\vert ^{2}-\left\vert -\xi+k_{2}\right\vert ^{2}}{\left\vert \xi -k_{2}\right\vert ^{2}+\left\vert-\xi -k_{2}\right\vert ^{2}}=2\frac{\sqrt{(1-\eta )(1+\eta )}t}{t^{2}(1+\eta )+(1-\eta)}.
\label{Asy in tunnel barrier}
\end{equation}

One obtains a perfect agreement between the perturbative scattering method and our multiband calculations for $|t^{\uparrow\downarrow}|^2$ and $\mathcal{A}$ (Fig.~2b). The transmission coefficient for an incoming propagative spin-$\uparrow$ electron into an outgoing propagative spin-$\downarrow$ electron is non-zero after SOI is branched on. The transmission \textit{vs.} incident kinetic energy and incident angle is different from the case of a simple exchange-step. The maximum of transmission depends also on the incidence angle or $t$ parameter. The $\mathbf{k}\cdot \mathbf{p}$ theory gives a maximum of asymmetry when the evanescent wavevector equals in magnitude the parallel incoming wavevectors in the CB.

\section{IV - Case of intrinsic Core SOI in the valence band: Chirality}

We now turn on the case of the VB of a tunnel junction composed of two \textit{p}-type ferromagnets separated by a thin tunnel barrier (3~nm in the present case). The barrier height have been chosen so as to match with the exchange strength ($0.3$~eV). The structure is free of any odd-potential $k$-terms ($\widehat{H}_D$=0) and only includes core SOI (\textit{p}-orbitals). Results are displayed in Fig.~1c for the transmission maps and Fig.~2c for the corresponding asymmetry resulting from a multiband $k\cdot p$ treatment. In the 2D-map calculation procedures obtained for a hole kinetic energy of $\epsilon=0.23$~eV, we have checked (Fig.~1c) that the whole numerical approaches (6, 14, 18 and 30-bands models) provide about exact similar data. The transmission scales within the range $15-45\times 10^{-3}$ with ($P^{^{\prime}}=0$ and $\Delta^{^{\prime}}=0$. Those results demonstrate that the absence of inversion symmetry ($T_d$) is not mandatory to observe an asymmetry $\mathcal{A}$. Fig.~2c displays the asymmetry $\mathcal{A}$ \textit{vs.} hole energy $\mathcal{E}$ for $k_\parallel=0.05$~nm$^{-1}$. The energy range covers the spin-$\uparrow,\downarrow$ heavy ($HH$) and light ($LH$)-hole subbands whereas the respective spin $\uparrow$ and $\downarrow$ split-off bands are not represented here. We refer to points (1) to (4) marked by vertical arrows in the following discussion. Here, the energy of the $HH\uparrow $ ($HH\downarrow $) corresponds to $0.15$~eV [$-0.15$~eV] as indicated by point (1) [(4)], the energy zero being taken at the top of the VB of the non-magnetic material. Correspondingly, one observes a large negative transmission asymmetry ($-60$\%) in this energy range for predominant majority spin $\uparrow$ injection as far as $HH\downarrow $ does not contribute to the current. At more negative energy [$\mathcal{E}<-0.15$~eV: point (4)], a sign change of $\mathcal{A}$ occurs at the onset of $HH\downarrow $ to reach about +20\%. From Ref.~\cite{dang2015} $\mathcal{A}$ changes sign two times at characteristic energy points corresponding to a sign change of the injected particle spin. Also, we have performed similar calculation for a simple contact~\cite{dang2015}. Remarkably, $\mathcal{A}$, although smaller, keeps the same trends as for the tunnel junction, except for a change of sign. Without tunnel junctions, $\mathcal{A}$ abruptly disappears as soon as $SO\downarrow $ contributes to tunneling \textit{i.e.}, when evanescent states disappear. In the case of tunnel junction, $\mathcal{A}$, although small, subsists in this energy range and this should be related to the evanescent character of the wavefunction in the barrier.

\section{V - Conclusions}

We have presented theoretical evidence for large interfacial tunneling asymmetry of carriers (scattering), electrons or holes, \textit{vs.} their incidence in exchange-split semiconductor structures. The effect of transmission asymmetry occurs in the CB via the SOI Dresselhaus interactions whereas intrinsic SOI of the \textit{p}-type VB is sufficient. This transmission asymmetry have been revealed by taking into account boundary wavefunctions matching, advanced multiband $\mathbf{k}\cdot \mathbf{p}$ calculations as well as scattering perturbation theory. After averaging over incoming states, a large surface current parallel to the barrier is results in an Anomalous Tunnel Hall effect.

\vspace{0.1in}

\begin{acknowledgments}
THD acknowledges Idex Paris-Saclay  and Triangle de la Physique for funding.
\end{acknowledgments}

\end {document}